stringar@science.unitn.it 
 \documentstyle[aps,prl,twocolumn,epsfig]{revtex}

 \draft

\begin{document}

 \title{ Scissors mode and superfluidity
 of a trapped Bose-Einstein condensed gas}

 \author{ D. Gu\'ery-Odelin and S. Stringari}
  \address{Dipartimento  di Fisica, Universit\`{a} di Trento,}
 \address{and Istituto Nazionale per la Fisica della Materia,
 I-38050 Povo, Italy}

  \date{\today}

  \maketitle

  \begin{abstract}
We investigate the oscillation of a dilute atomic gas generated by a sudden
rotation of the confining  trap (scissors mode).
This oscillation  reveals the 
effects of superfluidity  exhibited by a Bose-Einstein
condensate. The scissors mode is investigated also in a classical gas
above $T_c$ in various collisional regimes. The crucial difference 
with respect to the superfluid case arises from the occurrence of low
 frequency components, which are responsible for the rigid value of the 
moment of inertia. Different experimental procedures to excite 
the scissors mode are discussed.
   \end{abstract}

 \pacs{PACS numbers: 03.75.Fi, 05.30.Jp, 32.80.Pj}

 \narrowtext

 Superfluidity is one of the most spectacular consequences of
 Bose-Einstein condensation and has been the object of extensive
 experimental and theoretical work in the past, especially in connection
 with the physics of liquid helium \cite{helium}.
 Indirect signatures of superfluidity
 in trapped Bose-Einstein condensed gases
 are given by their dynamic behaviour
  at very low temperatures   \cite{mithd} which well confirms
 the predictions of the hydrodynamic theory of superfluids,
 as well as by the occurrence of
 spectacular interference phenomena \cite{mitc} which point out
 the importance of coherence
 effects, typical of superfluids. However, a direct evidence of
superfluidity
 is still missing in these systems.

 Important manifestations of superfluidity
 are associated with rotational phenomena. These include
  the strong reduction
 of the moment of inertia with respect to the classical rigid value
and the occurrence of quantized vortices 
\cite{donnelly}.
  These
 peculiar features are the direct consequence of the irrotational nature
 of the superfluid flow and have been already the object of 
theoretical
 work  also in the case of dilute trapped gases (see for example
\cite{mi} and references therein)

 The purpose of this work is to focus on the oscillatory behaviour
 exhibited by
 the rotation of the atomic cloud with respect to the
 symmetry axis of the  confining trap (see figure 1)
and on the corresponding superfluid
effects caused by Bose-Einstein condensation.
 A similar mode, called the scissors mode,
 is well known in nuclear physics \cite{nuclear}, where it
corresponds to the
out of phase
 rotation of the neutron and proton clouds, and its recent
systematic experimental investigation \cite{richter}
 has confirmed the occurrence of
 superfluidity in an important class of deformed nuclei.

 In the presence of a deformed external potential 
the restoring force  associated
 with the rotation of the cloud in the $x$-$y$ plane is proportional
 to the square of the
 deformation parameter $\epsilon$ of the trap  (see eq.(\ref{Vext}) below).
 The mass parameter is instead
 fixed by the moment of inertia. For a superfluid system this
 is given by the irrotational value and is hence  proportional to
$\epsilon^2$.
 As a consequence, the frequency of  the oscillation
 approaches a finite value when the
 deformation  tends to zero. Viceversa, in the absence
of
 superfluidity, the moment of inertia takes the rigid value and the
scissors
 mode exhibits low frequency components.

 Let us start our investigation by considering a Bose-Einstein condensed
 gas at equilibrium in  a deformed  potential of the
 form
 \begin{equation}
 V_{\rm ext}({\bf r}) = \frac{m}{2}\omega_x^2 x^2+\frac{m}{2}
 \omega^2_yy^2+\frac{m}{2}\omega^2_z z^2
 \label{Vext}
 \end{equation}
with $\omega^2_x= \omega^2_0(1+\epsilon)$ and 
$\omega^2_y= \omega^2_0(1-\epsilon)$
where $\epsilon$ gives the deformation of the
trap
 in the $x$-$y$ plane.
 For large enough samples one can safely use the Thomas-Fermi approximation
 for the ground state density
 \begin{equation}
 n_0({\bf r})=(\mu-V_{\rm ext}({\bf r}))/g
 \label{TF}
 \end{equation}
 where the strength parameter $g$ is related to the scattering
 length $a$ by the relation $g=4\pi\hbar^2a/m$.
We consider a gas initially  at
equilibrium
 in the $(x',y',z)$ frame.
At $t=0$ one rotates abruptly the eigenaxis of
the trap from
their initial position to $(x,y,z)$ by a small angle
-$\theta_0$ (see figure 1).
   As a consequence of the sudden
 rotation, the system will be no longer  in equilibrium and will
 start oscillating. If the   angle $\theta_0$ is not too large the
 oscillation will correspond to a rotation of the cloud in the $x$-$y$
plane (scissors mode).
 The equations describing this rotation  
 can be easily obtained at zero temperature starting from the time
 dependent Gross-Pitaevskii equation for the order parameter which,
in the Thomas-Fermi regime, takes the typical form of
the hydrodynamic equations of
superfluids \cite{rmp}:
 \begin{eqnarray}
 \frac{\partial n}{\partial t} & + & \mbox{div}(n{\bf v})=0
 \label{continuity}\\
 m\frac{\partial {\bf v}}{\partial t} & + & \mbox{\boldmath $\nabla$}_{\bf
r}\bigg(
 V_{\rm ext}({\bf r})+gn+\frac{mv^2}{2}\bigg)=0
 \label{euler}
 \end{eqnarray}
Starting from (\ref{continuity},\ref{euler}) one easily obtains, in the linear regime, 
the closed set of equations 
 \begin{eqnarray}
 \frac{d}{dt}\langle xy\rangle & = & \langle xv_y+yv_x\rangle
 \label{hdss1}
 \\
 \frac{d}{dt}\langle xv_y+yv_x\rangle & = & -2\omega_0^2\langle xy\rangle
 \label{hdss2}
 \end{eqnarray}
involving the relevant quadrupole variable $\langle xy\rangle = 
\int d{\bf r}xy n({\bf r},t)/N$. The  corresponding density profile 
can be parametrized in the
form
 \begin{eqnarray}
 n({\bf r},t) & = & (\mu - V_{\rm ext}({\bf r}) -m\omega^2_0\alpha(t) xy)/g
 \label{n(t)}
 \end{eqnarray}
where $\alpha$ is a small time dependent coefficient related to 
$\langle xy\rangle$ by 
 \begin{equation}
 \langle xy\rangle
 =-{2\omega_0^2\alpha(t)\over 3m \omega^2_x\omega^2_y}
\langle V_{\rm ext}\rangle_0
 \label{xy}
 \end{equation}
and
$\langle V_{\rm ext}\rangle_0$ is the expectation value of the harmonic
potential
 at equilibrium. The associated 
velocity field is irrotational and has the form
${\bf v}({\bf r},t) \propto \mbox{\boldmath $\nabla$}(xy)$.
It is worth noticing that only if $\alpha \ll \epsilon$ will the
parametrization (\ref{n(t)}) describe a rotation of the sample with
the angle fixed by the relation $\alpha(t)=2\epsilon\theta(t)$. If
instead $\alpha$ is larger than $\epsilon$, then (\ref{n(t)})
corresponds to a traditional quadrupole deformation characterized 
by a change of the intrinsic shape
and the connection
with the geometry of the scissors is lost. 
In order to obtain
a visible signal of rotational type in the density (\ref{n(t)})
 the deformation parameter
should not be consequently too small.  
Assuming $\alpha \ll \epsilon$ (or,
equivalently, $\theta \ll 1$) one finds, from (\ref{hdss1}), (\ref{hdss2}) and (\ref{xy}),
the equation
 \begin{equation}
 { d^2\theta(t)\over dt^2} +2\omega_0^2\theta = 0
 \label{thetat}
 \end{equation}
The solution corresponding to 
the chosen initial conditions
 $\theta(0)=\theta_0$ and $\theta'(0)=0$ is 
 $\theta=\theta_0\cos(\omega t)$ and oscillates with
frequency   $\omega=\sqrt{2}\omega_0$.

 Let us now consider the behaviour of the system in the absence of
Bose-Einstein condensation.
The simplest case is the high $T$, classical regime where
 analytic solutions are available in the framework
 of the Boltzmann kinetic equations. We will use  the method of the
 averages, already employed to discuss the damping of the quadrupole
oscillation \cite{enstn}. 
The method provides  the following  closed set of equations
 \begin{eqnarray}
 \frac{d}{dt}\langle xy\rangle & = & \langle xv_y+yv_x\rangle
 \label{a}
 \\
 \frac{d}{dt}\langle xv_y-yv_x\rangle & = & 2\epsilon\omega_0^2\langle
xy\rangle
 \label{b}
 \\
 \frac{d}{dt}\langle xv_y+yv_x\rangle & = & 2(\langle v_xv_y\rangle-
 \omega_0^2\langle xy\rangle)
 \label{c}
 \\
 \frac{d}{dt}\langle v_xv_y\rangle & = & -
 \omega_0^2\langle xv_y+yv_x\rangle-\epsilon\omega_0^2\langle
xv_y-yv_x\rangle
 \nonumber\\
 & & -\frac{\langle v_xv_y\rangle}{\tau}
 \label{d}
 \end{eqnarray}
where the averages are taken here 
in both coordinate and velocity space and the
collisional term has been evaluated in the linear regime using a
 gaussian approximation for the distribution function \cite{enstn}.
The relaxation time entering (\ref{d}) is fixed by the relation
$\tau =5/(4\gamma_{\rm coll})$  where
$\gamma_{\rm coll} =n(0)v_{\rm th}\sigma/2 $ is the classical
collisional rate with $n(0)$ the central density
of the atomic cloud, $v_{\rm th}=\sqrt{8k_BT/\pi m}$ the thermal velocity
and
$\sigma =8\pi a^2$ the elastic cross-section. Notice that collisions affect
only the equation for the variable $\langle v_xv_y\rangle$. In fact the
other variables are conserved by
the elastic collisions \cite{huang}.
It is also worth noticing that the angular momentum
 $m(xv_y-yv_x)$ is not a constant of motion, due to the absence of
symmetry
 in the confining potential and is  coupled, through eq.(\ref{b}),  to the
quadrupole  variable $\langle xy \rangle$.
 It is finally interesting to note that
eqs. (\ref{a}-{\ref{d}), with the collisional term set equal to zero, 
exactly hold also in the case of a non interacting Bose or  Fermi gas.
In this case the effects of quantum statistics enter 
 only through the  initial conditions of the 
corresponding dynamic variables.
For oscillations of small
amplitude the density profile of the classical gas corresponds to the
gaussian
parametrization
\begin{equation}
 n({\bf r},t)\;\propto\; e^{-(V_{\rm ext}({\bf
r})+\alpha(t)m\omega_0^2xy)/kT}
 \label{nclassical}
 \end{equation}
and the expectation value of $\langle xy\rangle$
is given by the same
relation (\ref{xy}) previously derived for the $T=0$ Bose-Einstein condensed
gas. The initial
conditions corresponding to  the sudden rotation of the
gas are given by
$\langle xy\rangle_{t=0} = \epsilon
\theta_0\langle x^2+y^2\rangle_{t=0}$,
 $\langle xv_y\pm yv_y\rangle_{t=0} = 0$
and  $\langle v_xv_y  \rangle_{t=0} =0$. The vanishing of
$\langle xv_y\pm yv_y\rangle_{t=0}$ follows from the absence of
currents at $t=0$, while the vanishing of 
$\langle v_xv_y  \rangle_{t=0}$ is the consequence of 
the initial isotropy of the velocity distribution.  
-
By using the relationship $\alpha(t)=-2\epsilon\theta(t)$
holding for small rotational angles the  equations of motion 
(\ref{a}-\ref{b})
can be usefully rewritten in the form
 \begin{equation}
 \bigg(
 \frac{d^4\theta}{dt^4}+4\omega^2_0\frac{d^2\theta}{dt^2}
 +4\epsilon^2\omega^4_0\theta\bigg)
 +\frac{1}{\tau}\bigg(\frac{d^3\theta}{dt^3}+2\omega^2_0\frac{d\theta}{dt}
 \bigg)= 0
 \label{theta4t}
 \end{equation}
and the initial conditions take the form $\theta(0)=\theta_0$,
$\theta''(0)=-2\omega_0^2\theta(0)$ and
$\theta'(0)=\theta'''(0)= 0$.
Notice that eq.(\ref{theta4t}) differs from the corresponding equation 
(\ref{thetat}) for the superfluid regime.
It gives rise to different solutions
propagating at high and low frequency. For small
values
of the deformation parameter $\epsilon$, the former can be identified
 with the $\ell_z=2$ irrotational quadrupole oscillation, i.e. the classical 
counterpart of the superfluid oscillation described by (\ref{thetat}). The
latter instead corresponds to the rotational mode of the
system and is absent in the superfluid case. 
Actually, if $\epsilon$ becomes too small, the variable $\theta$ looses
its geometrical meaning. In this case  the physical variables are
on the one hand  the
 moment $\langle xy \rangle$, which characterizes
the high lying quadrupole mode and is coupled 
with the variables $\langle xv_y + yv_y\rangle$ and 
$\langle v_xv_y\rangle$, and on the other hand 
  the angular momentum
$m\langle xv_y-yv_x\rangle$ which  becomes a constant of motion 
(see eq.(\ref{b})).

Let us discuss the different collisional regimes predicted by
(\ref{theta4t}).
 In the collisionless regime (first term in
(\ref{theta4t})) the two frequencies are undamped and given, respectively,
by $\mid\omega_x\pm\omega_y\mid$. In the opposite, hydrodynamic
limit (second term) only the high-lying oscillation survives
with frequency $\omega=\sqrt2\omega_0$, while the low-lying solution becomes
overdamped as $\tau \to 0$, in agreement with the diffusive nature
of the transverse waves predicted by classical hydrodynamics \cite{lfm}. 
A maximum
damping is obtained for $\omega_0\tau\sim 1$ a condition easily achievable 
in current experiments.

 The chosen initial condition, corresponding to a sudden rotation
of the sample, gives rise to the excitation of both the low and high frequency
modes.
The resulting
time evolution of the observable $\theta(t)$ is shown in fig.2 and 3
for two different collisional regimes and the  choice
$\epsilon=0.3$.
In fig.2 we have made the ``collisionless"
choice $\epsilon\omega_0\tau \sim 6$, so that both the high lying and low
lying modes have small damping. The figure clearly shows the combined
signals
propagating with frequencies $\mid\omega_x\pm\omega_y\mid$ respectively.
In fig.3 we have instead made the
choice $\epsilon\omega_0\tau \sim 0.15$ corresponding
to large damping. With such a choice
the low frequency rotational mode is overdamped and the remaining
oscillation exhibited by the curve is due to the high  frequency component
The achievement of the overdamped, ``hydrodynamic" 
regime for the low frequency
mode, is favoured by the choice of small values of $\epsilon$.

  From the comparison between the curves of fig. 2 and 3 it emerges very
clearly
 that the main feature of the superfluid regime (dashed line in fig.2 and fig.3)
is the absence of
 low frequency components. This behaviour
should be easily verificable
experimentally.

Untill now, we have discussed the superfluid case ($T=0$) and
the classical regime $T > T_c$. Below $T_c$ the system can be described
in terms of a two-fluid model
and  one  needs two different angles to describe the motion
of the system. 
Neglecting interaction effects between the two fluids 
the superfluid component would be governed by (\ref{thetat}) 
while the thermal part would evolve   according to (\ref{theta4t}). 
Of course in this case the relaxation time  $\tau$ should be evaluated by
taking into account
Bose statistics.
Inclusion of interaction effects 
between the condensate and the thermal component
would lead to a damping of the  
superfluid oscillation, similarly to 
what happens 
in the case of the quadrupole mode
\cite{jin}.

 The sudden rotation of the trap is not the only
 way to excite the scissors mode. In the last part of this letter
 we discuss an alternative procedure
 which further emphasizes
 the superfluid nature of the  Bose-Einstein condensed gas. 
We consider a gas initially in equlibrium within  
a trap  rotating with  frequency $\Omega$.
 Experimental techniques to achieve rotating configurations of this type
 are in progress \cite{ens}.
 At the time $t=0$
we suddenly stop the rotation of the trap and the gas, due to
its
 inertia, will start rotating.
 In the absence of superfluidity the moment
 of inertia  is large and the oscillations will be characterized
by
 low frequency components and large amplitudes. Viceversa, if the system
 is superfluid the moment of inertia is small (proportional to
 $\epsilon^2$) and  the oscillations will be characterized by large
  frequencies and  small amplitudes.
 The equations of motion  describing the rotation are still given by 
(\ref{hdss1}) and (\ref{hdss2}) but the initial conditions are different, corresponding 
to the presence of a current term in the system at $t=0$. One finds
$\theta(0)=0$ and $\theta'(0)=\Omega$.
In the classical
case one has the additional conditions  $\theta''(0)=\theta'''(0)=0$.
The amplitude of the oscillations, in the
collisionless regime, scale 
as $\Omega/\omega_0$ for the superfluid and as
 $\Omega/(\omega_0\epsilon)$ for the classical gas. 
The difference is due to the fact that at $t=0$ the velocity field
generated by the rotation is very different in the two cases.
In the superfluid it is given by the irrotational 
form ${\bf v}=-\Omega\epsilon\mbox{\boldmath $\nabla$} (xy)$, while in the classical case
by the rotational form ${\bf v}=\mbox{\boldmath $\Omega$} \times{\bf r}$.
If the 
amplitude of the oscillation becomes too
large the dynamic variable $\langle xy \rangle$
cannot be longer simply connected with the angle $\theta$. This easily happens
in the classical case where the amplitude of the oscillation is magnified by
the factor
$1/\epsilon$.

The excitation of the scissors mode based on a sudden switching off of the
rotation of the trap would allow also for  an explicit determination of the
moment of inertia $\Theta$ of the gas defined by the 
linear relation
$m\langle xv_y-yv_x\rangle_{t=0} =\Omega\Theta/N$. 
In fact the knowledge of the
time evolution of $\langle xy\rangle$
permits to calculate directly the value of the angular momentum 
 at $t=0$. From eq.(\ref{b}),
which holds for classical as well as for quantum systems, 
one has $m\langle xv_y-yv_x\rangle_{t=0} =-
2\epsilon\omega_0^2\int d\omega F(\omega)/\omega$ with
$F(\omega)$ defined by
\begin{equation}
\langle xy \rangle(t)=\int d\omega F(\omega)(e^{i\omega t}-e^{-i\omega
t})/2i\;.
\end{equation}
On the other hand, from the model independent equation (\ref{a}),
one easily finds the result $\langle xv_y+yv_x \rangle_{t=0}=\int
d\omega F(\omega)\omega$ so that, using 
the relation 
$\langle xv_y+yv_x \rangle_{t=0} = 
\Omega\langle x^2-y^2\rangle_{t=0}$ predicted by  linear 
response theory \cite{note}, one finally obtains
the useful expression 
\begin{equation}
\Theta=\Theta_{\rm rig} (\omega^2_y-\omega_x^2){\langle x^2 -y^2\rangle_{t=0}
\over \langle x^2 +y^2\rangle_{t=0}}
\frac{\int d\omega F(\omega)/\omega}{\int d\omega F(\omega)\omega}
\label{F}
\end{equation}
relating the moment of inertia of the system to the measurable Fourier
signal $F(\omega)$. In this equation $\Theta_{\rm rig}= Nm
\langle x^2+y^2\rangle$ is the rigid value of the moment of inertia.
Notice that eq.(\ref{F}) provides an exact relationship for the moment of
inertia holding for classical 
as well as Bose or Fermi interacting systems confined by a harmonic trap
of the type (\ref{Vext}). If the system oscillates
with the single frequency $\omega=\sqrt 2\omega_0$, as happens in the $T=0$
Thomas-Fermi superfluid regime, one immediately finds the result
$\Theta=\epsilon^2\Theta_{\rm rig}$
corresponding to the
irrotational value of the moment of inertia. Viceversa, in a classical gas
or in a normal Fermi gas trapped by a harmonic potential, 
the response is dominated by low frequency components,
of order $\epsilon\omega_0$,
and eq.(\ref{F}) yields the rigid value for the moment of inertia.

 We acknowledge fruitful discussions with J. Dalibard, L. Pitaevskii and F. Zambelli.
 This project was  supported by
 INFM through the Advanced
 Research Project on BEC and by MURST.

\begin{figure}
\begin{center}
\epsfig{file=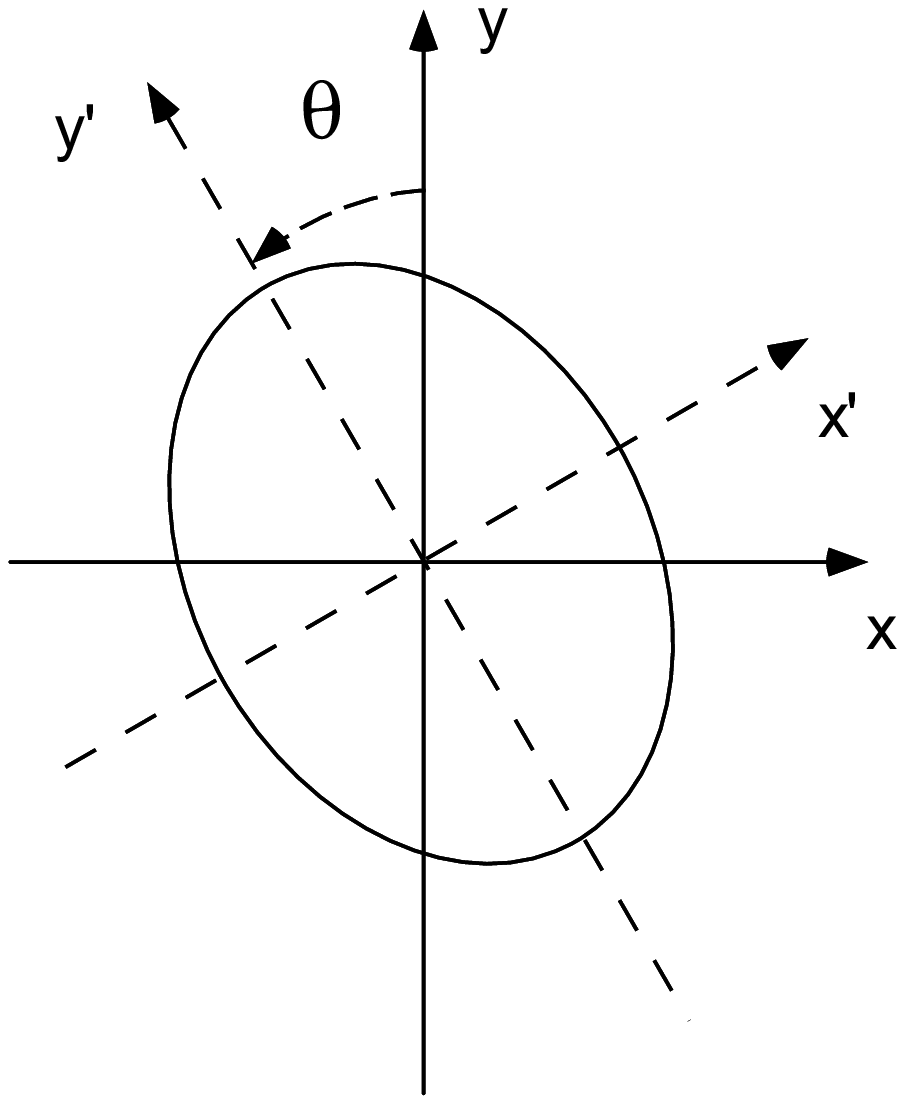, width=4.5cm,angle=0}
\begin{caption}
{Exciting the scissor mode : initially the gas is  thermalized in an
anisotropic trap. One then abruptly rotates the eigenaxis of the trap by a small angle.}
\end{caption}
\end{center}
\label{fig1}
\end{figure}

\begin{figure}
\begin{center}
\epsfig{file=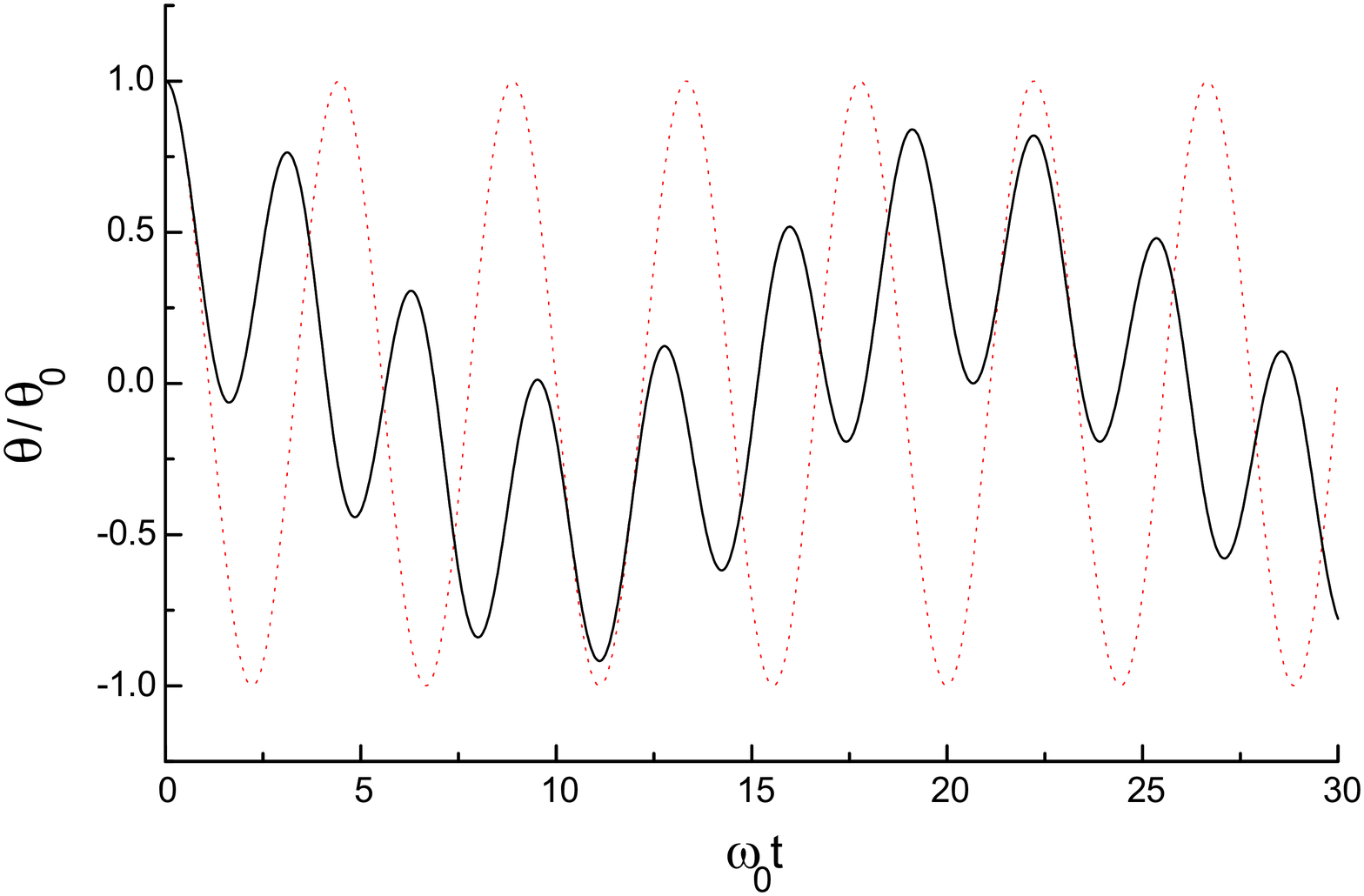, width=7cm,angle=0}
\begin{caption}
{Angle $\theta$ as a function of time for a classical gas (full line) in the ``collisionless"
regime ($\epsilon\omega_0\tau=6$ and $\epsilon=0.3$) and for a superfluid (dashed line).}
\end{caption}
\end{center}
\label{cl}
\end{figure}

\begin{figure}
\begin{center}
\epsfig{file=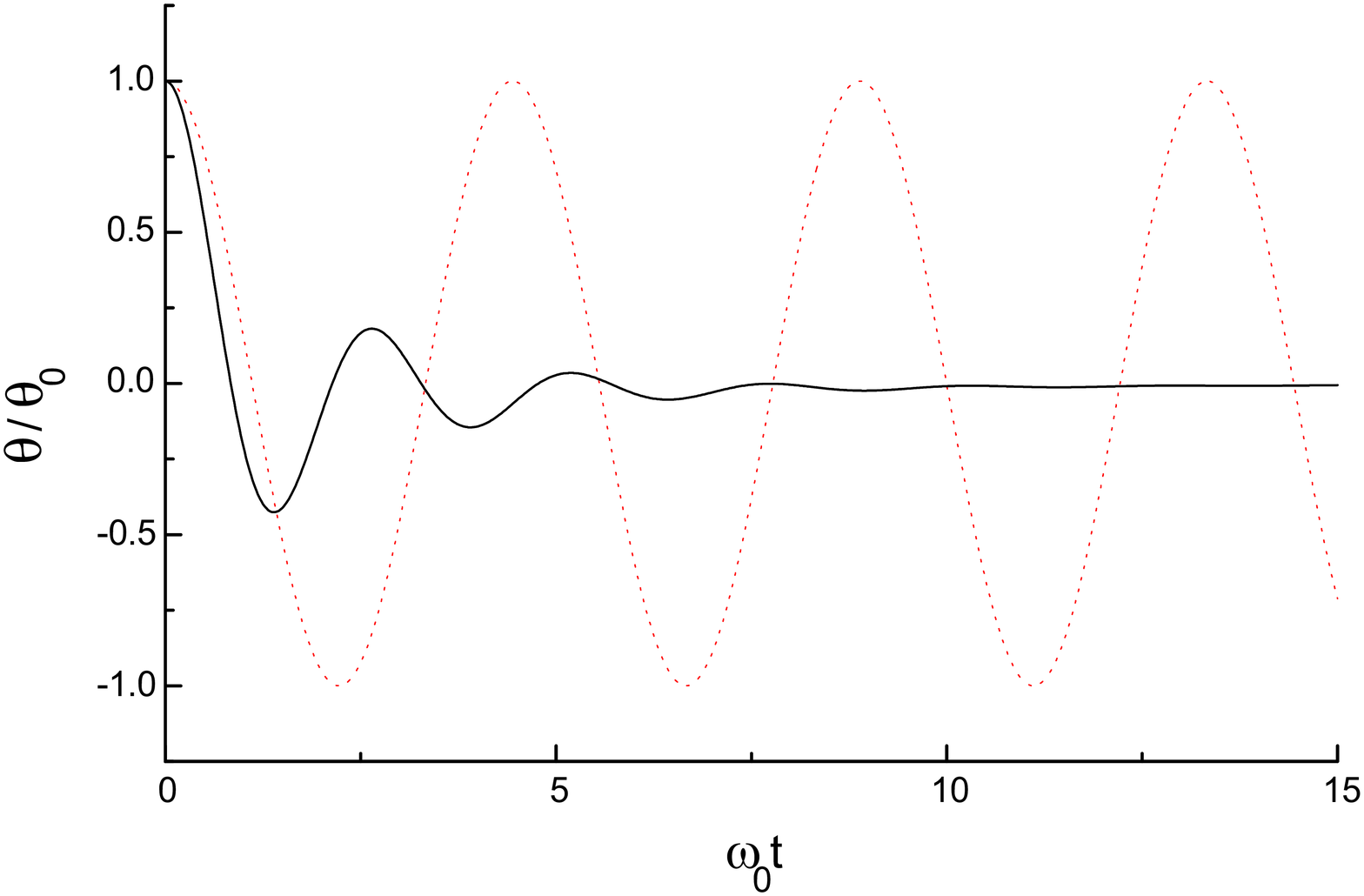, width=7cm,angle=0}
\begin{caption}
{Angle $\theta$ as a function of time for a classical gas (full line) in the ``hydrodynamic"
regime ($\epsilon\omega_0\tau=0.15$ and $\epsilon=0.3$) and for a
superfluid (dashed line).}
\end{caption}
\end{center}
\label{fig3}
\end{figure}


\begin{references}
\bibitem{note} 
According to linear response theory, the constraint $-\Omega
\sum_i\ell_z^{(i)}$,
yields the result $\langle xv_y+yv_x \rangle
=\Omega \Xi$ where 
$\Xi= \sum_{nq} (\Pi_n-\Pi_q) \langle n|
\sum_i (x^{(i)}v_y^{(i)} + y^{(i)}v^{(i)}_x) |q\rangle
\langle q | \sum_i\ell_z^{(i)} | n\rangle / (E_q-E_n)$
is the static response function
relative to the corresponding operators with $\Pi_n=e^{-E_n/k_BT}/\sum_ne^{-E_n/k_BT}$.
 Using the model independent 
commutation relations $[H, \sum_i x^{(i)} y^{(i)}] =
 -i\hbar\sum_i(x^{(i)}v^{(i)}_y+y^{(i)}v_x^{(i)})$
and $[\ell_z, xy] = im\hbar(x^2-y^2)$ as well as the completeness relation, 
one immediately derives the relationship 
$\langle xv_y+yv_x \rangle= 
\Omega\langle x^2-y^2\rangle$.
\end{references}
 \end{document}